\documentclass[journal]{IEEEtran}
\usepackage[square,sort&compress,comma,numbers]{natbib}
\usepackage{color}
\usepackage{graphicx}
\usepackage{amsfonts}
\usepackage{epstopdf}
\usepackage{latexsym}
\usepackage{amssymb}
\usepackage{amsmath}
\usepackage{fixltx2e}
\usepackage{algorithm,algorithmic}
\usepackage{multicol}
\usepackage{multirow}
\usepackage[T1]{fontenc}
\usepackage{threeparttable}
\usepackage{mathrsfs}

\ifCLASSOPTIONcompsoc
	\usepackage[caption=false, font=normalsize, labelfont=sf, textfont=sf]{subfig}
	\else
	\usepackage[caption=false, font=footnotesize]{subfig}
\fi

\begin{document}

\title{An Energy-efficient Compressed Sensing Based Encryption Scheme for Wireless Neural Recording}

\author{Xilin~Liu,~\IEEEmembership{Member,~IEEE,}
        Andrew~G. Richardson,~\IEEEmembership{Senior Member,~IEEE} \\
        and~Jan~Van der Spiegel,~\IEEEmembership{Life Fellow,~IEEE}

\thanks{Xilin Liu and Jan Van der Spiegel are with the Department of Electrical and Systems Engineering (ESE), University of Pennsylvania, Philadelphia, PA, 19104 USA. E-mail: xilinliu@seas.upenn.edu.}
\thanks{Andrew~G. Richardson is with the Department of Neurosurgery, University of Pennsylvania, Philadelphia, PA, 19104 USA.}
\thanks{}
\thanks{Digital Object Identifier}
}

\maketitle

\begin{abstract}
This paper presents a compressed sensing (CS) based encryption scheme for wireless neural recording. An ultra-high efficiency was achieved by leveraging CS for simultaneous data compression and encryption. CS enables sub-Nyquist sampling of neural signals by taking advantage of their intrinsic sparsity, while the CS process simultaneously encrypts the data with the sampling matrix being the cryptographic key. To share the key over an insecure wireless channel, we implemented an elliptic-curve cryptography (ECC) based key exchanging protocol. Local key shuffle and updating were adopted to eliminate the risks of potential information leakage. CS was executed in an application-specific integrated circuits (ASIC) fabricated in 180nm CMOS technology. Mixed-signal circuits were designed to optimize the power efficiency of the matrix-vector multiplication (MVM) of the CS operation. The ECC was implemented in a low-power Cortex-M0 based microcontroller (MCU). To be protected from timing attacks, the implementation avoided possible data-dependent branches. A wireless neural recorder prototype has been developed to demonstrate the proposed scheme. The prototype achieved an 8x data rate reduction and a 35x power saving compared with conventional implementation. The overall power consumption of ASIC and MCU was 442$\mu$W during the encrypted wireless transmission. The average correlated coefficient between the reconstructed signals and the uncompressed signals was 0.973, while the ciphertext-only attacks (CoA) achieved no better than 0.054 over 200,000 attacks. This work demonstrates a promising data compression and encryption scheme that can be used in a wide range of low-power signal recording systems with security requirements.
\end{abstract}

\begin{IEEEkeywords}
Hardware security, compressed sensing, cryptographic circuits, low power, mixed-signal IC, wireless, neural recording.
\end{IEEEkeywords}

\maketitle

\vspace{5mm}

\section{Introduction}

\IEEEPARstart{L}arge-scale neural recording with high energy efficiency and safety is crucial to the growing number of therapies employing closed-loop neurostimulation \cite{Sun2014} and neuroprosthetics \cite{Bouton2016} to treat brain injury and disease. Although the circuits and system community has devoted a considerable amount of effort to improve the performance and power efficiency of neural recording systems, few investigations have been done to mitigate the security risks. In fact, cybersecurity issues have already emerged in FDA-approved medical devices \cite{FDA2019}. Medical devices, including neural interfacing devices, pose serious risks from malicious attacks. Compromised neural interfacing devices may not only disclose critical health-related information, but also leave the users vulnerable to life-threatening attacks. Thus, it is urgent to investigate secure neurotechnologies.

Battery-powered wireless neural recording devices are especially vulnerable to malicious attacks, and their restrained energy budget imposes a significant challenge to implementing data encryption. A typical wireless neural recorder consists of the following key blocks: low-noise amplifiers and filters, analog-to-digital converters (ADCs), wireless transmitters, and optional digital signal processing units. Energy-efficient circuit design techniques of each block have been extensively discussed in the literature. For a low-power neural amplifier design with a noise efficiency factor (NEF) of 3, the energy cost is around 0.1nJ/bit \cite{Harrison2003}. For a low-power ADC design with a Walden figure-of-merit (FoM) of 100fJ/conv-step, the energy cost is around 0.01nJ/bit \cite{Murmann2015}. For a low-power wireless transceiver design for biomedical applications, an energy cost of 1nJ/bit is typical, while ultra-low power designs with energy costs below 1nJ/bit have also been reported \cite{Teng2017,Rosenthal2019,Lee2020}. However, the hardware implementation of data encryption standards by application-specific integrated circuits (ASICs) and general-purpose processors typically takes 1nJ/bit and 10nJ/bit, respectively \cite{Alioto2019,Han2015,Devlin2013}. As a result, standard encryption algorithms may not meet the power requirement for direct integration into low-power neural recorders without optimization.

\begin{figure}[!ht]
  \centering
  \includegraphics[width=0.48\textwidth]{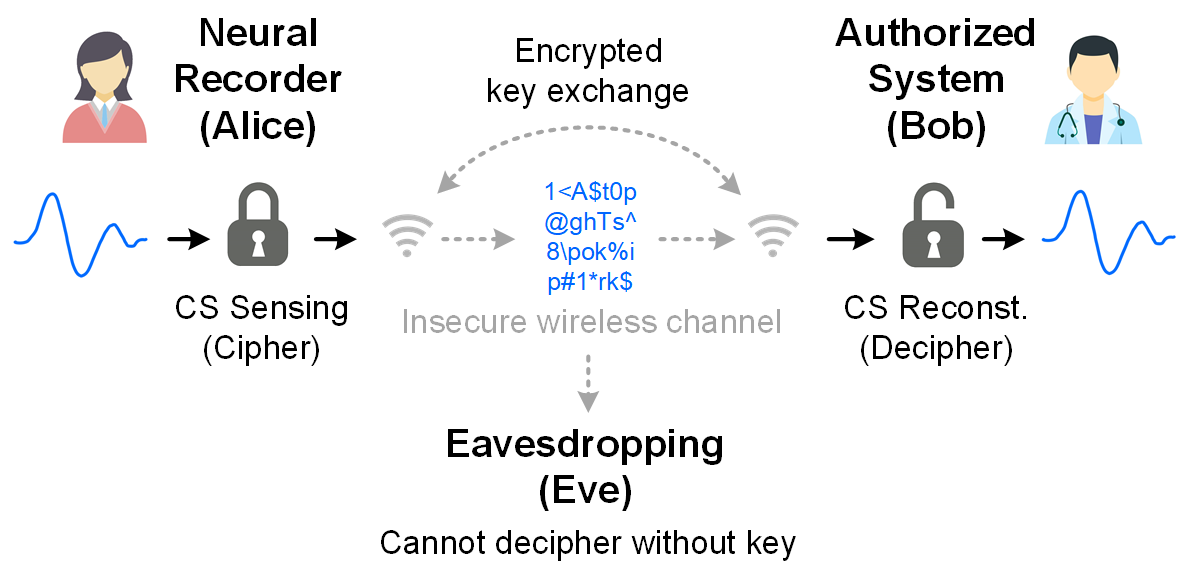}\\
  \caption{Illustration of the proposed neural recording system with CS based encryption. The conventional character Alice represents the neural recorder, Bob represents the authorized external system, and Eve represents the illegitimate parties that tend to steal private information by eavesdropping. }\label{sys_concept}
\end{figure}

In this work, we proposed a novel encryption scheme for neural recording that achieves ultra-high energy efficiency by leveraging compressed sensing (CS) technique, as illustrated in Fig. \ref{sys_concept}.
The key concept of CS is that a sparse signal can be sampled at a reduced rate (below Nyquist frequency) based on the actual amount of information it contains \cite{Donoho2006}. Neural signals are proven to be sparse in certain domains and pre-learned dictionaries \cite{Aviyente2007,Tao2018,Liu2016TBioCAS}. Recent studies have successfully demonstrated highly efficient CS based neural recorder designs \cite{Tao2018,Liu2016TBioCAS,Liu2015ISCAS,Zhao2018,Aprile2018,Mangia2020,Shoaran2014}. Moreover, the CS theory also permits its application in data encryption \cite{Rachlin2008}. It has been proven that CS can provide a computational guarantee of secrecy, provided that an adversary doesn't know the sampling matrix \cite{Zhang2016}. Recently, several works have explored this property in image processing \cite{Zhang2018Security,Cho2020} and internet of things (IoT) applications \cite{Mangia2019,Chen2019TVLSI}. We presented in this paper, to the best of our knowledge, the first implementation in neural recording systems.

For CS based encryption to be successfully implemented in a neural recording system, several challenges must be addressed. First, the sampling matrix (i.e. the cryptographic key) must be safely exchanged between the neural recorder and authorized external system. Second, CS sampling is a linear projection process, where the energy features of the signal may be revealed without an accurate decipher \cite{Kaushik2019}. Interception of energy features could be used to disclose age, gender, and potentially other information about the subject, thus, additional mechanisms must be introduced to protect these features. Third, the hardware implementation must be protected from side-channel attacks, such as timing attacks \cite{Coron1999,Kocher1996}. Last but not least, the overall encryption cannot lead to a significant power penalty. To overcome these challenges, we proposed a novel system that combines an optimized integration of an ASIC and a general-purpose microcontroller (MCU). The ASIC performs mixed-signal CS operations at ultra-low power consumption, while the MCU handles the low duty-cycle key sharing, shuffling, and updating. An elliptic-curve cryptography (ECC) based protocol was implemented in the MCU with time-constant executions. A prototype design using the proposed scheme achieved an 8x data rate reduction and a 35x power saving compared with traditional implementation.

The rest of this paper is organized as follows. Section II describes the operating principles of the proposed system. Section III presents the implementation details. Section IV shows the experimental results. Section V discusses the limitations and future directions. Finally, Section VI concludes the paper.

\section{Operating Principles}

\subsection{CS for Joint Signal Compression \& Encryption} \label{sec2_cs}
We first briefly review the fundamentals of CS. Suppose the input signal $\textbf{x}$ has a sparse representation $\textbf{s}$ on a certain basis ${\bf \Psi}$. CS theory predicts that $\textbf{x}$ can be sampled at a reduced rate (depending on its sparsity) with nearly no information loss. The compressed measurement $\textbf{y}$ can be expressed as:
\begin{equation}
\textbf{y} = {\bf \Phi x}\label{eq_cs}
\end{equation}
where $\textbf{x} \in \mathbb{R}^{N\times 1}$, $\textbf{y} \in \mathbb{R}^{M\times 1}$, and ${\bf \Phi} \in \mathbb{R}^{M\times N}$. Note that $N$$>$$M$, and the term $N/M$ is referred to as compression ratio (CR). Although $\textbf{y}$ cannot be solved directly from Eq. (\ref{eq_cs}), if the sampling matrix ${\bf \Phi}$ is incoherent with ${\bf \Psi}$ (obeying the restricted isometry property (RIP) \cite{Baraniuk2007,Wakin2008,Candes2005}), the sparse representation $\textbf{s}$, thus the original signal $\textbf{x}$, can be solved as a convex optimization problem \cite{Wakin2008}:
\begin{equation}
\text{min} \: ||\textbf{s}||_{0 \; (or \; 1)}, \;\; \text{s.t.} \;\;  {\bf y = \Phi x = \Phi \Psi^{-1} s}
\end{equation}
In this work, we used a $\mathscr{l}$1-norm based reconstruction algorithm \cite{Donoho2006}. It has been proven that a binary random matrix ${\bf \Phi}$ consisting of 0 and 1 meets the minimum requirements in fulfilling the incoherent requirement \cite{Wakin2008}. Prior works showed improved reconstruction performance and resistance to noises by having additional resolution \cite{Liu2016TBioCAS, Liu2015ISCAS}. Here, we adopted a 4-bit ${\bf \Phi}$ consisting of elements of $\{$0, $\pm$1/8, $\pm$2/8, $\pm$3/8, $\pm$4/8, $\pm$5/8, $\pm$6/8, $\pm$7/8$\}$ following a Gaussian distribution. The hardware implementation will be discussed in Section \ref{hw_asic}.

Since the introduction of CS in 2006 \cite{Donoho2006}, many algorithms and techniques have been proposed for improving the reconstruction performance. In addition to the convex optimization based approaches (under different norm regularizations), greedy strategy approximation based algorithms (e.g. orthogonal matching pursuit \cite{Zhang2011}), dictionary learning \cite{Zhang2015}, adaptive CS \cite{Mangia2020Adapted}, as well as deep learning \cite{Sun2017} have been proposed to speed up the process of finding the optimal solution. Our objective in this work was not to achieve record-breaking reconstruction performance. Rather, we focused on the hardware design and optimization of the sampling end (i.e. the neural recorder).

The secrecy property of CS has also been rigorously discussed in the literature \cite{Rachlin2008,Mayiami2013,Zhang2016,Yang2014}. Although achieving Shannon's perfect secrecy is conditional \cite{Mayiami2013}, computational secrecy can be guaranteed \cite{Rachlin2008}. Encryption algorithms with computational secrecy are commonly adopted in cryptography standards, given that extracting information without the key is a nondeterministic polynomial time problem (NP-problem) \cite{Zhang2016,Yang2014}. However, since CS sampling is a linear projection process, the energy of \textbf{x} can be revealed in \textbf{y} without accurately deciphering the measurement \cite{Kaushik2019}. The energy features of neural signals can contain biometrics and private information of the subject. The information may be leaked to over-the-air eavesdroppers if no additional protection is used.

To mitigate this risk, Chen and colleagues proposed a method of inserting watermarks to mask the energy features \cite{Chen2019TVLSI}. Cambareri and colleagues proposed a multiclass encrypting scheme \cite{Cambareri2015}. In our application of neural recording, we hope not to degrade the reconstructed signal quality or significantly increase the hardware complexity. Thus, we proposed a pseudo-random key shuffle as well as a synchronized key updating for disturbing the energy features. The power penalty of this scheme was negligible in the overall system due to its low active duty cycle.

\subsection{Elliptic-Curve Cryptography and Key Exchanging} \label{sec2_ecc}

There are two types of encryption schemes, known as symmetric encryption and asymmetric encryption \cite{Simmons1979}. Symmetric encryption uses one key to cipher and decipher the messages. Asymmetric encryption uses a pair of keys: a public key to cipher the messages, and a private key to decipher them. In symmetric encryption, the key must be kept secret once shared between the sender and receiver. Conversely, in asymmetric encryption, the public keys are available to all, and the private keys are never shared. 
Commonly used symmetric encryption standards include DES, AES, and RC4/5/6, and popular asymmetric encryption standards include RSA, DSA, and ECC \cite{Alioto2019}.

Although asymmetric encryption algorithms have the advantage of stronger security, it is often at the expense of high computational costs. As a result, people often use asymmetric encryption to share the key, and then use the key for symmetric encryption. Here, we adopted a similar strategy. CS is a symmetric encryption strategy since the key (the sampling matrix) is used in both ciphering and deciphering the messages. This key should be shared using an asymmetric encryption scheme.

Among established asymmetric encryption schemes, RSA is most widely used. However, recent ECC based encryption can achieve the same level of security as RSA but with a shorter key length, a lower computation cost, and a lower latency. For example, a 224-bit ECC achieves an equivalent security level of a 2048-bit RSA, which was a security level recommended by the National Institute of Standards and Technology (NIST) in 2015 \cite{NIST2015}. In this work, we adopted a 256-bit ECC based key exchanging protocol, namely Elliptic-curve Diffie-Hellman (ECDH) \cite{Bernstein2006}. ECDH is an ECC variant of the classic Diffie-Hellman protocol \cite{Diffie1976}. ECDH allows two parties to establish a shared secret key independently. The shared secret key can then be used directly or for deriving other keys, which in our case are the CS sampling matrices. The detailed implementation is described in Section \ref{hw_ECDH}.

\subsection{Framework of the Proposed Hybrid Encryption}

In a typical scenario of neural signal transmission, the neural recorder (the conventional character Alice) sends the sampled data to the authorized external system (Bob) via a low-power insecure wireless channel. Illegitimate parties (Eve) may steal the messages by eavesdropping. In this work, we adopted a commonly used threat model that Eve knows the encryption algorithms (including the parameters of elliptic curves, field, etc.), the communication protocol, as well as the public keys, but doesn't have access to the unciphered plaintext and the private key (the CS sampling matrices ${\bf {\Phi_S}}$). This is also known as a ciphertex-only attack (CoA) model \cite{Chen2019TVLSI}. Considering the practical use scenarios of wearable or implantable neural recording devices, we assumed that the adversary won't gain physical ownership of the device during its signal transmission, but may non-invasively detect the power profile of the devices (e.g. using electromagnetic approaches) and deduce timing characteristics from power analysis \cite{Mulder2005}. Finally, the attack range we considered in this work is within a personal area network (PAN), not telecommunication networks.

\begin{figure}[!ht]
  \centering
  \includegraphics[width=0.44\textwidth]{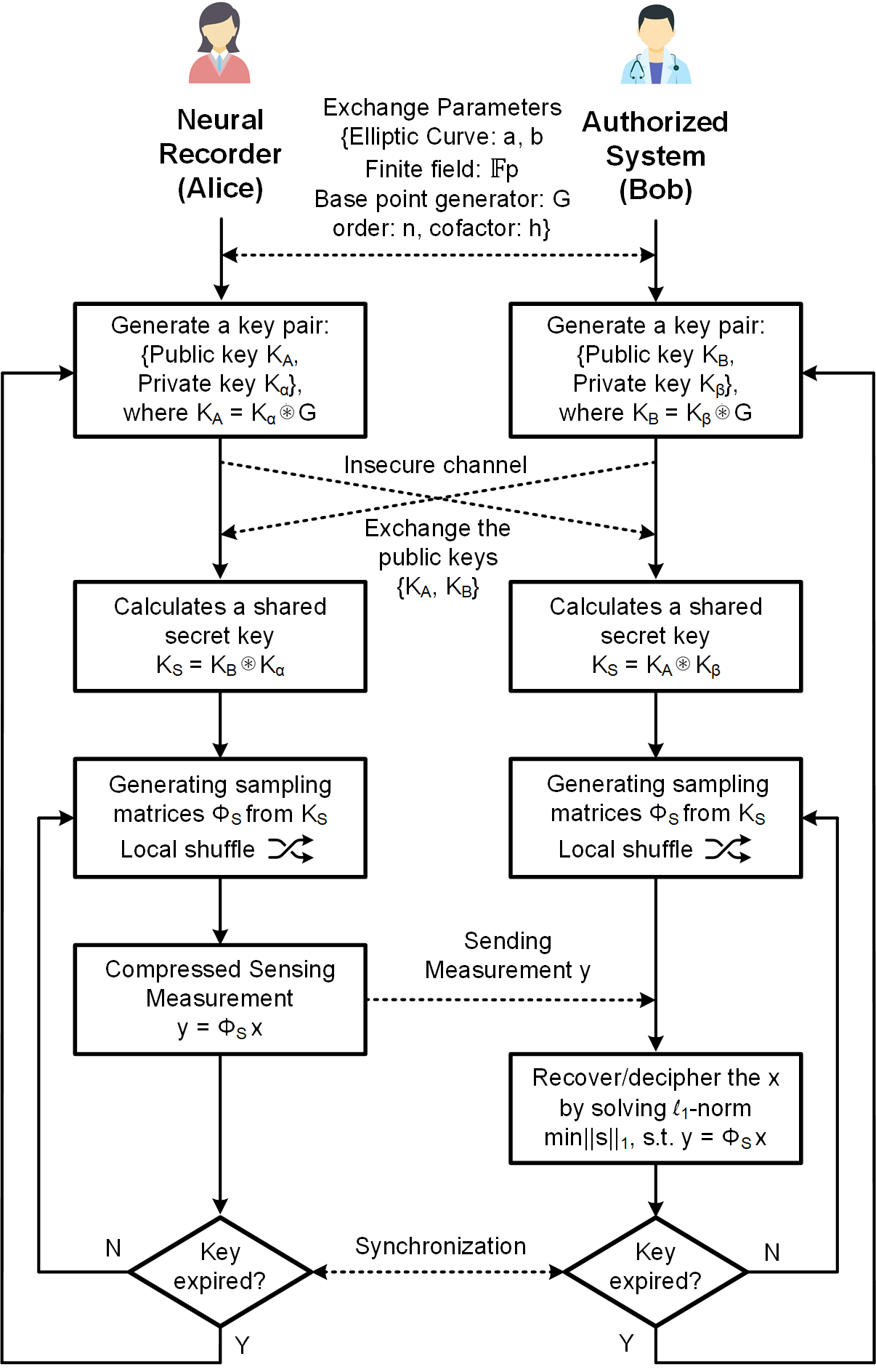}\\
  \caption{The operation of the proposed neural recording system. Dashed lines indicate the communication is via an insecure wireless communication channel.}\label{sys_protocol}
\end{figure}

Fig. \ref{sys_protocol} illustrates the basic operation principles of the proposed cryptographic neural recording system. The operation procedure is as follows.
\begin{enumerate}
  \item Alice and Bob first agree on a set of domain parameters (public) for the cryptography, including the parameters of the elliptic curve, the generator point ${\bf G}$, etc.;
  \item Alice picks a private key ${\bf K_\alpha}$ and generates a public key ${\bf K_A = K_\alpha \circledast G}$, where $\circledast$ is a multiplication defined by ECC; Similarly, Bob picks a private key ${\bf K_\beta}$ and generates a public key ${\bf K_B = K_\beta \circledast G}$;
  \item Alice and Bob exchange their public keys ${\bf K_A}$ and ${\bf K_B}$;
  \item Alice and Bob generate a shared secret key ${\bf K_s}$ using their own private keys and the public keys provided by the other party: 
    \begin{align}
    {\bf K_s} &= {\bf K_A \circledast K_\beta = (K_\alpha \circledast G) \circledast K_\beta} \; \; (\leftarrow Bob) \\
    & = {\bf K_\alpha \circledast (G \circledast K_\beta) = K_\alpha \circledast K_B} \; \; (\leftarrow Alice)
    \end{align}
    so that Alice and Bob have the same secret key, but Eve cannot get it;
  \item Alice and Bob individually generate a set of sampling matrices ${\bf {\Phi_S}}$ using the secret key ${\bf K_s}$;
  \item Alice performs CS on acquired neural signal \textbf{x}, and sends the lower-dimensional measurement \textbf{y} to Bob;
  \item Bob recovers the neural signal \textbf{x} from \textbf{y} by solving optimization problem using ${\bf {\Phi_S}}$;
  \item Alice and Bob shuffle the ${\bf {\Phi_S}}$ according to a pre-agreed protocol, and then repeat the CS encryption;
  \item To prevent from using the same set of ${\bf \Phi_S}$ repeatedly, Alice and Bob would update the ${\bf K_S}$ (thus the sampling matrix ${\bf \Phi_S}$) periodically on a synchronized manner.
\end{enumerate}

It should be noticed that the encryption scheme implemented in this work doesn't verify identities during the public key exchanging process. The authentication process can be established in various ways, such as implementing the digital signature algorithm \cite{Johnson2001}.

\section{System Implementation}

\begin{figure}[!ht]
  \centering
  \includegraphics[width=0.4\textwidth]{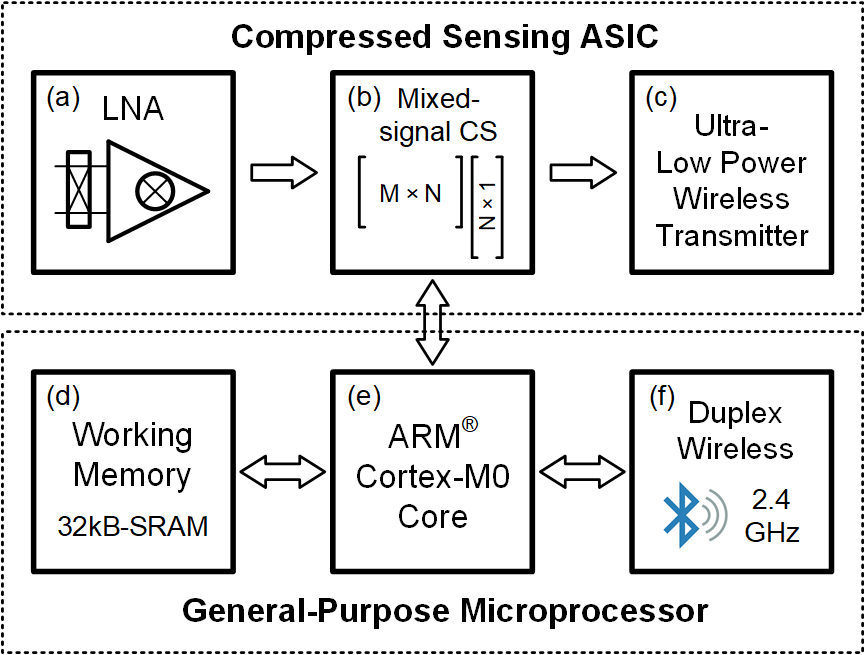}\\
  \caption{The high-level block diagram of the proposed system. The system mainly consists of an ultra-low power ASIC for CS (always ON), and a general-purpose Cortex-M0 MCU for ECC based key sharing (low duty cycle). }\label{sys_diagram}
\end{figure}

The high-level block diagram of the proposed neural recording system (Alice) is shown in Fig. \ref{sys_diagram}. The system mainly consists of an ultra-low power ASIC and a general-purpose MCU. The ASIC executes CS measurements of neural signals using mixed-signal circuits and sends out the measurements using an on-chip wireless transmitter (Tx). The MCU executes the ECC based key exchanging and handshakes with external receivers via a 2.4GHz duplex wireless transceiver (Tx + Rx) for PAN communication. The ASIC design and MCU implementation are discussed in the subsequent sections.

On the other hand, a computer interfacing device (Bob) has been designed. This device integrates a MCU and corresponding wireless transceivers for pairing with the neural recorder (Alice). A standard USB 2.0 port is integrated for high-speed communication with the computer system. A MATLAB based user interface has been developed for device configuration and data logging \cite{PennBMBI2015,PennBMBI2014ISCAS}.

\subsection{ASIC Design for Mixed-Signal CS}\label{hw_asic}

\begin{figure}[!ht]
  \centering
  \includegraphics[width=0.49\textwidth]{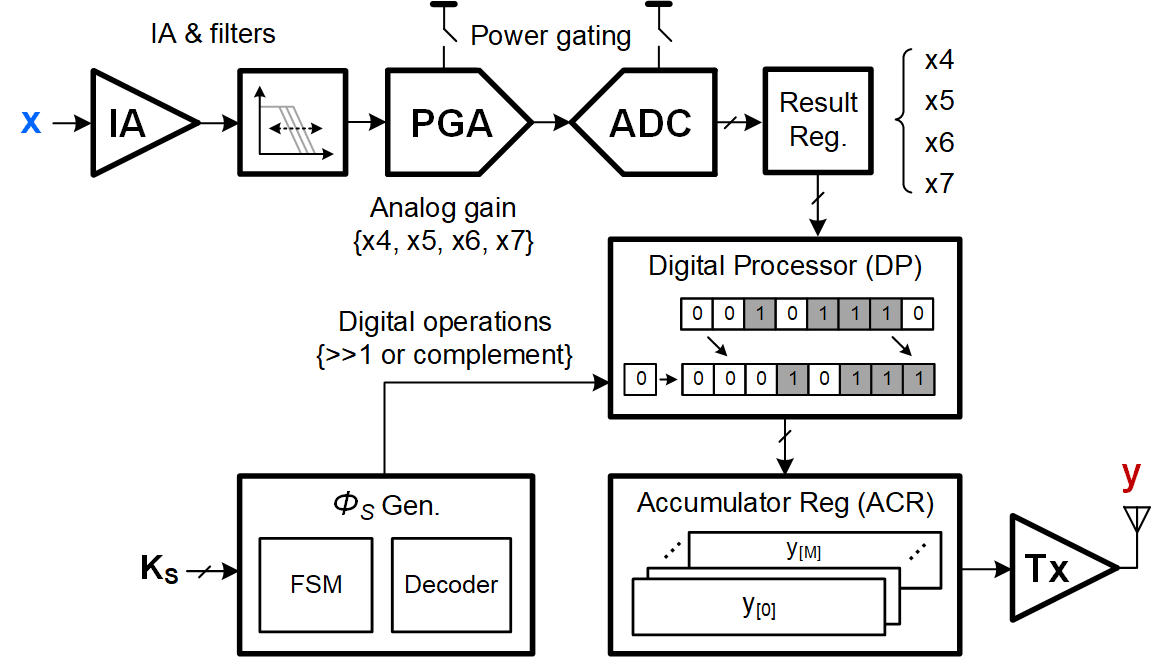}\\
  \caption{The block diagram of the mixed-signal ASIC for CS operation. }\label{diagram_asic}
\end{figure}

The block diagram of the ASIC design is shown in Fig. \ref{diagram_asic}. The ASIC integrates low-noise instrumentation amplifiers (IA) and filters, a programmable gain amplifier (PGA), a successive-approximation register (SAR) ADC, a CS processor, an ultra-low power wireless transmitter, and peripheral circuits including power management units (not shown in the figure). 16-channel IA and filters were integrated for pairing with microelectrode array (MEA), but only one recording channel is used in this work. The IA and wireless transmitter design reuses aspects of our previous work \cite{ISCAS2017,Sensors2020,PID2016}.

The ASIC design focused on improving the energy efficiency of the CS operation. In particular, the repeated matrix-vector multiplication (MVM) between the input signal $\textbf{x}$ and the $\bf \Phi$ dominates the system's power consumption. In this work, we avoided the power and silicon area consuming digital multiplication by using a combination of analog processing and simple digital logic. As discussed in the Section \ref{sec2_cs}, we adopted $\bf \Phi$ with a resolution of 4-bit. The common factor of 1/8 among $\{$0, $\pm$1/8, $\pm$2/8, $\pm$3/8, $\pm$4/8, $\pm$5/8, $\pm$6/8, $\pm$7/8$\}$ is combined with the IA gain. Analog gain values of $\{\times$4, $\times$5, $\times$6, $\times$7$\}$ are provided by a programmable gain amplifier (PGA) before digitization. Results of $\times$2 and $\times$3 are generated by shifting the samples of $\times$4 and $\times$6 after digitization by 1-bit to the right ($>>$1), respectively. In this way, power hungry digital multiplication is replaced by simple logic operations. Negative numbers are generated by digital complementation. Table \ref{mixed_signal_multi} summarizes the operations. It should be noticed that $\times$1 was sampled directly bypassing the PGA by default, but it can also be generated by shifting the samples of $\times$4 by 2 bits.

\begin{table}[!ht]
\centering
\caption{Mixed-signal multiplication of $\textbf{x}$ and $\bf \Phi$} \label{mixed_signal_multi}
\begin{threeparttable}
\begin{tabular}{|c|c|c|c|}
\hline
\multirow{2}{*}{} & Analog   & \multicolumn{2}{c|}{Digitial} \\ \cline{2-4}
                & PGA gain      & Bit shift      & Complement       \\ \hline
0               & - $^{\dag}$   & -              & -                \\ \hline
$\pm$ 1/8 $^*$  & -             & -              & - or (-1)      \\ \hline
$\pm$ 2/8       & $\times$ 4    & $>>$ 1          & - or (-1)      \\ \hline
$\pm$ 3/8       & $\times$ 6    & $>>$ 1          & - or (-1)      \\ \hline
$\pm$ 4/8       & $\times$ 4    & -              & - or (-1)      \\ \hline
$\pm$ 5/8       & $\times$ 5    & -              & - or (-1)      \\ \hline
$\pm$ 6/8       & $\times$ 6    & -              & - or (-1)      \\ \hline
$\pm$ 7/8       & $\times$ 7    & -              & - or (-1)      \\ \hline
\end{tabular}
    \begin{tablenotes}
       \footnotesize
       \item $^{\dag}$  '-' indicates no operation required.\\
       \item $^*$ The common factor of 1/8 is combined with the IA gain.
     \end{tablenotes}
     \end{threeparttable}
\end{table}

\begin{figure}[!ht]
  \centering
  \includegraphics[width=0.45\textwidth]{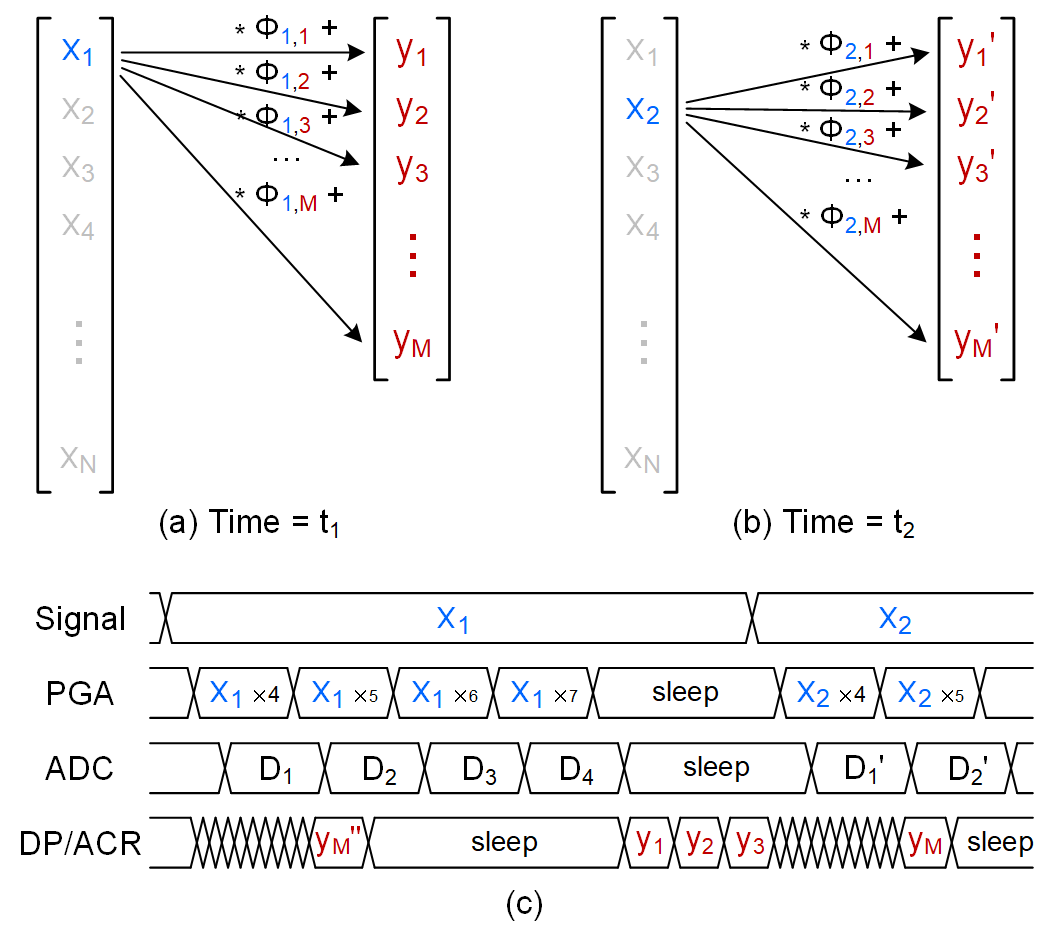}\\
  \caption{Illustration of the mixed-signal MVM operation and the timing of the CS measurement. (a) and (b) show the arithmetical operation at time point $t_1$ and $t_2$, respectively. (c) shows the timing diagram of the analog sampling, digitization, and the digital processing. }\label{cs_eq}
\end{figure}

The signal processing flow of the CS measurement is as follows. The neural signal is first amplified and conditioned by the low-noise IA and filters. During an input period of $t_i$ (Fig. \ref{cs_eq}), the signal $x_i$ is sampled four times in a sequence with PGA gain values of $\{\times$4, $\times$5, $\times$6, $\times$7$\}$. The four samples are digitized by the SAR ADC and saved in corresponding registers. The digital processor (DP) processes the samples based on the $\Phi_{i,j}$, where $j \in (1,M)$. The $M$ results are sent to the accumulator registers (ACR) in 16 bits. The ACR adds $N$ samples during one CS measurement (eq. \ref{eq_cs}). 

The input data rate $f_x$ is determined by the bandwidth of the target neural signal. The output data rate $f_y$ is 1/CR of $f_x$. Given the frequency nature of intracranial EEG signals, the $f_x$ is typically less than 500 Hz \cite{Niedermeyer2004}. The PGA consists of an operational amplifier with an open-loop gain of 75dB over PVT in simulation. The closed-loop gain of the PGA is set by programmable feedback resistors. The layout of the resistors was carefully placed for good unit matching. Post-layout Monte-Carlo simulation results showed that the PGA achieved the required linearity for the 9-bit digitization. The SAR ADC was designed with a 10-bit resolution with an effective number of bits (ENoB) better than 9-bit. The PGA and SAR ADC were designed with a bandwidth of at least 4$f_x$ for the proposed CS operation, and the DP processes the data in $M f_x$. In practice, the bandwidth of the analog blocks was designed with margin, and these blocks were gated when not activated for power saving. $N$ and $M$ are programmable on-chip for a CR of 2x to 16x. $M$ is programmable to be 64, 96, or 128. The 16-bit CS measurement \textbf{y} is sent off-chip serially. All CS operations are in real-time.

\subsection{MCU Implementation of Key Exchanging} \label{hw_ECDH}

As discussed in Section \ref{sec2_ecc}, ECC based algorithms have advantages over conventional asymmetric cryptography algorithms in terms of speed, security level (given a key length), as well as the corresponding computational costs. These features make it attractive for both security-critical applications (e.g. virtual currency \cite{Turan2019}) and resource-constrained applications, including wireless neural recording.

Among established elliptic curves, we chose Curve25519 for our application, because of its low requirements in memory and computational resources. Curve25519 and the corresponding Diffie-Hellman functions were originally proposed by Daniel Bernstein in 2006 \cite{Bernstein2006}. The function is a field-restricted scalar multiplication on an elliptic curve $E$:
\begin{equation}
y^2 = x^3 + 486662x^2 + x \; \; \; \; (x, y) \in \mathbb{F}_{p}^2
\end{equation}
where $p$ is $2^{255}-19$. When a point $P$ (on the curve $E$) multiplies a scalar $S$, it adds to itself ($S$-1) times to a point Q, which remains on the curve $E$ (the set forms an abelian group). The computation only uses the $x$-coordinate, thus is also called $x$-coordinate scalar multiplication. The $x$-coordinate scalar multiplication is repeated twice (on each party) in the ECDH protocol for generating the public key and the shared secret key (as illustrated in Fig. \ref{sys_protocol}), respectively.

In this work, we adopted a 256-bit key using a radix $2^{32}$ representation for the code implementation. The $x$-coordinate scalar multiplication can be efficiently computed using the classic Montgomery ladder \cite{Montgomrey1987}. Algorithm \ref{alg1} describes the operation in pseudo-codes. Each $Ladderstep$ performs one differential addition and one doubling \cite{Sasdrich2014}.

\begin{algorithm}
\caption{{\small Scalar Multiplication (Original)}}
\small
\label{alg1}
\begin{algorithmic}[1]
\renewcommand{\algorithmicrequire}{\textbf{Inputs:}}
\renewcommand{\algorithmicensure}{\textbf{Output:}}
  \REQUIRE $P$ (a point on the curve E), $S$ (a scalar)
  \ENSURE $Q$ (a point on the curve E)
  \STATE $Q \Leftarrow Initial$ $point$
  \FOR {$each$ $bit$ $b$ $of$ $S$ ($254$ $downto$ $0$)}
  \IF {$b$ $is$ $1$}     \STATE {$swap$ $the$ $values$ $of$ $P$ $and$ $Q$} \ENDIF
  \STATE $(P,Q)$ $\Leftarrow$ $Ladderstep(P,Q)$
  \IF {$b$ $is$ $1$}     \STATE {$swap$ $the$ $values$ $of$ $P$ $and$ $Q$} \ENDIF
  \ENDFOR
\end{algorithmic}
\end{algorithm}

\begin{algorithm}
\caption{{\small Scalar Multiplication (Time-Constant Implementation)}}
\small
\label{alg2}
\begin{algorithmic}[1]
\renewcommand{\algorithmicrequire}{\textbf{Inputs:}}
\renewcommand{\algorithmicensure}{\textbf{Output:}}
  \REQUIRE $P$ (a point on the curve E), $S$ (a scalar)
  \ENSURE $Q$ (a point on the curve E)
  \STATE $Q \Leftarrow Initial$ $point$
  \FOR {$each$ $bit$ $b$ $of$ $S$ ($254$ $downto$ $0$)}
  \IF {$b$ $is$ $0$}
    \STATE $(P,Q)$ $\Leftarrow$ $Ladderstep0(P,Q)$
  \ELSE
    \STATE $(P,Q)$ $\Leftarrow$ $Ladderstep1(P,Q)$
  \ENDIF
\ENDFOR
\end{algorithmic}
\end{algorithm}

In order to make the implementation immune to timing attacks, all input-dependent branches or operations, such as the conditional swap in the original algorithm, should be avoided. In this work, we modified the $Ladderstep$ function into two functions $Ladderstep0$ and $Ladderstep1$, as described in Algorithm \ref{alg2}. These two functions have identical timing. The execution of either function depends on the loop's variable bit $b$; thus, the timing dependence of the input data is eliminated. In addition, the initial coordinates were randomly projected in each execution according to \cite{Coron1999} for resistance to differential power attacks (DPA).

The 256-bit multiplication and squaring are the most computationally intensive operations. The 32-bit Cortex-M0 executes 32-bit multiplication in a single clock cycle, however, the returned results are in 32-bit instead of a full 64-bit. The 256-bit multiplication was implemented as a three-level Karatsuba multiplication with time-constant implementation \cite{Karatsuba1962,Dull2015}. Squaring operations use the same Karatsuba algorithm, but at a faster computing speed, thanks to the arithmetic simplification and memory access reduction \cite{Nascimento2015}.

\section{Experimental Results}

\begin{figure}[!ht]
  \centering
  \includegraphics[width=0.4\textwidth]{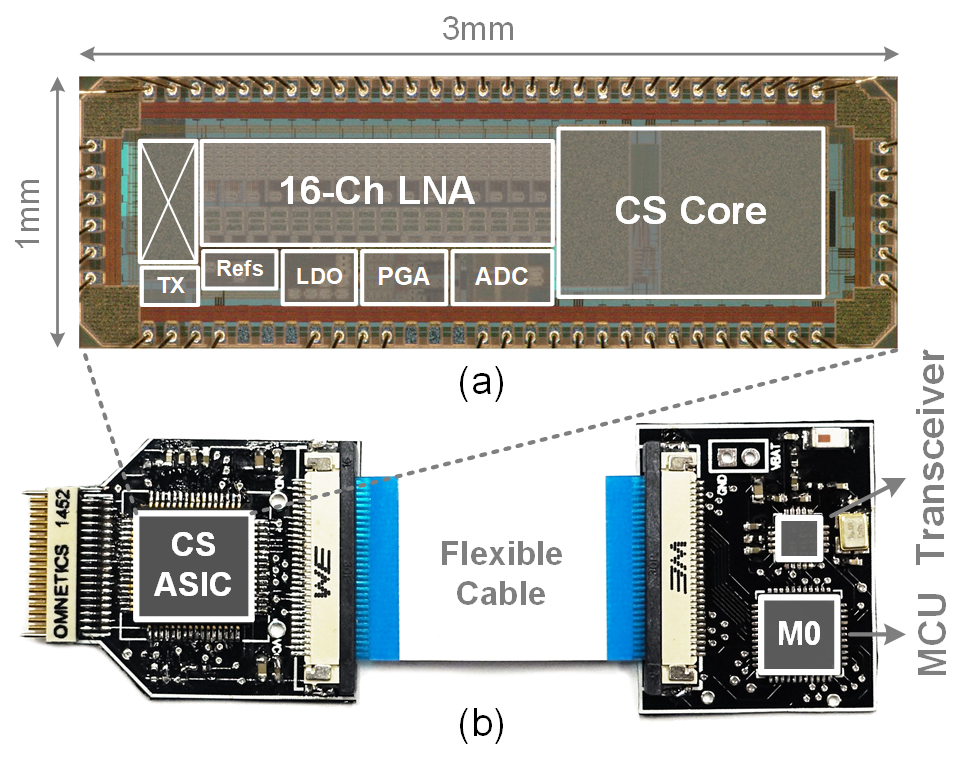}\\
  \caption{(a) A micrograph of the fabricated CS ASIC. (b) A photo of the assembled neural recording device. The device consists of a main PCB integrating the ASIC and an extension PCB integrating the MCU and a wireless transceiver. }\label{photo_device}
\end{figure}

The ASIC was fabricated in standard 180nm CMOS technology, occupying a silicon area of 2.5mm$\times$0.6mm excluding the IO pads (Fig. \ref{photo_device} (a)). The system was assembled on a 4-layer printed circuit board (PCB) (Fig. \ref{photo_device} (b)).
The main PCB contained the ASIC and a micro-connector for pairing with MEA. The MCU and the 2.4GHz wireless transceiver were assembled on an extension PCB, which was connected to the main PCB via a flexible cable. The device was powered by 3.7V lithium batteries. On-chip low-dropout regulators (LDOs) provide 1.8V analog and digital supplies to the ASIC, while the MCU and wireless transceiver use a 3.3V supply provided by an external LDO on board. The weight of the assembled device was 4.7g including a 46mAh battery.

The device was fully tested for functionality and performance. The measured noise of the IA was 2.31$\mu$V with an integral bandwidth of 0.5 to 250Hz. The IA gain was programmable from 40 to 54dB. The measured distortion at 100Hz was below -60dB, and the common-mode rejection ratio was above 73dB. The bandwidth of the PGA was 20kHz. The measured ENoB of the SAR ADC was 9.3 bit.

\begin{figure}[!ht]
  \centering
  \includegraphics[width=0.5\textwidth]{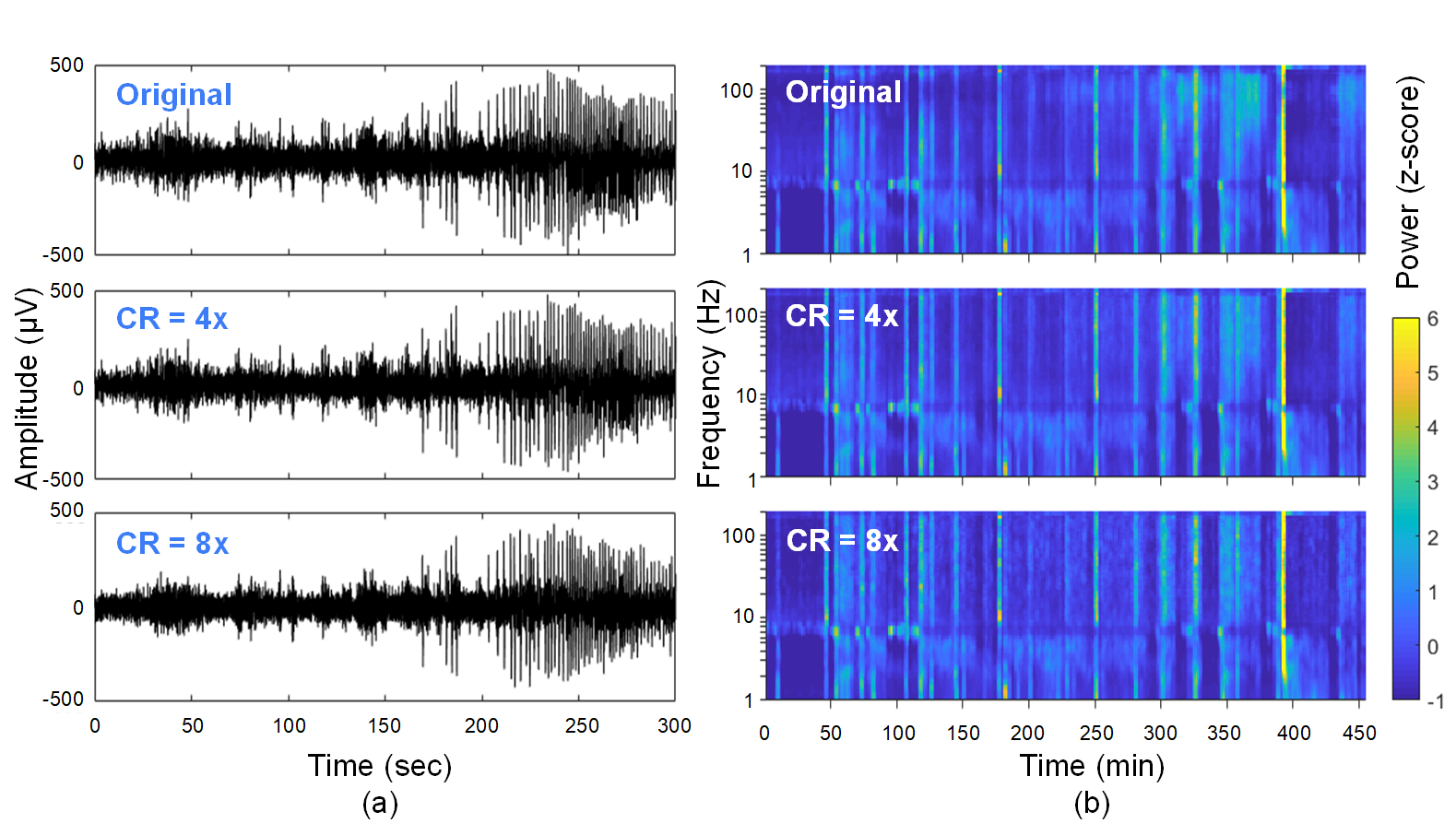}\\
  \caption{Experimental results of CS and signal reconstruction with different compression ratio (CR). (a) A 10-min segment of signal during the onset of a seizure event. (b) A 7.5-hour segment of spectrogram showing multiple seizure events. The three rows show the uncompressed signal, CR = 4x, and CR = 8x, respectively. }\label{exp_seizure}
\end{figure}

The CS function was tested using pre-recorded intracranial EEGs of epilepsy patients \cite{Wagenaar2015}. The signal was carefully reviewed and the seizure onset times were annotated by experts. In our experiment, we used a subset of the database that contains the recordings of two patients. The recorded EEG was replayed by an arbitrary signal generator in a resolution of 16-bit, followed by a 5th order low-pass filter with a cut-off frequency of 250Hz. A $\mathscr{l}$1-norm based reconstruction algorithm was implemented in MATLAB \cite{Donoho2006,Liu2016TBioCAS}. Fig. \ref{exp_seizure} shows the experimental results using one of the patients' data.
The reconstructed signals from a CR of 4x and 8x are plotted in comparison with the original signal without compression. $M$ was fixed at 128 in this experiment. The time-domain waveforms are shown in Fig. \ref{exp_seizure} (a). The spectrograms of a continuous recording of 7.5 hours are shown in Fig. \ref{exp_seizure} (b). The computed PSNR is 32.75dB at a CR of 8x. The resulting loss due to compression is below the thermal noise floor of intracranial EEG recording \cite{Niedermeyer2004}, indicating a sufficient performance for research and clinical use.

\begin{figure}[!ht]
  \centering
  \includegraphics[width=0.495\textwidth]{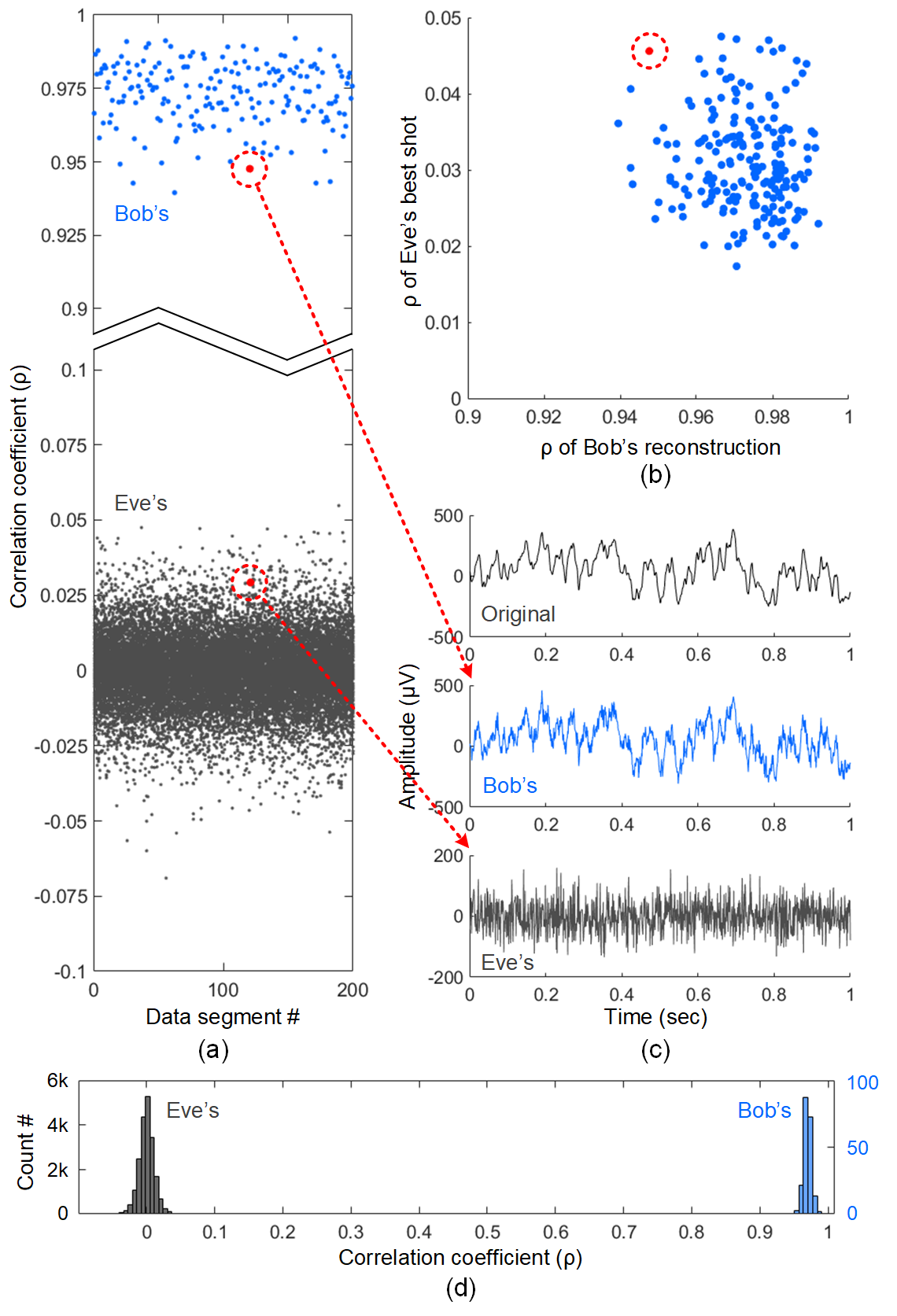}\\
  \caption{Experiment of 200,000 mock attacks using the CoA model. (a) The correlation coefficients ($\rho$) between Bob's and Eve's reconstructed signals and the original signals. (b) The scatter graph of Bob's reconstruction vs. Eve's best shot. (c) The time-domain waveforms of the trails where the $\Delta \rho$ between Bob's and Eve's reconstruction was minimum. (d) Histograms of the $\rho$ of Bob's and Eve's reconstructions. }\label{exp_attack}
\end{figure}

We tested the neural recording system under mock attacks using the CoA model. 
Fig. \ref{exp_attack} shows the results of a total of 200,000 CoA attacks to 200 data segments randomly selected from the two patients' recordings.
For each data segment, Bob had one reconstruction using the genuine key, while Eve made 1000 reconstruction attempts using randomly generated keys. Here we assumed Eve had prior knowledge of the targeting signals' characteristics, thus Eve used the same basis $\Psi$ as Bob for the signal reconstruction. Fig. \ref{exp_attack} (a) shows the correlation coefficients $\rho$ (the higher the better) between Bob's and Eve's reconstructed signals and the original signals. The $\rho$ as defined by Pearson was calculated as:
\begin{equation}
\rho = \frac{N \displaystyle \sum_{i=1}^{N} x_i \hat{x_i} - \sum_{i=1}^{N}x_i \sum_{i=1}^{N}\hat{x_i} }{\sqrt{ \displaystyle N\sum_{i=1}^{N}x_i^2 - (\sum_{i=1}^{N} x_i)^2} \sqrt{ \displaystyle N\sum_{i=1}^{N}\hat{x_i}^2 - (\sum_{i=1}^{N} \hat{x_i})^2}}
\end{equation}
where ${\bf \hat{x}}$ is the reconstructed signal, $N$ is the dimension of the data segment. The $\rho$ of Eve's reconstructions are within a random noise level. Fig. \ref{exp_attack} (b) is the scatter graph of the results from the 200 data segments with $x$-coordinate being the $\rho$ of Bob's reconstruction and $y$-coordinate being the $\rho$ of Eve's best shot. It should be noted that Eve doesn't know which one is the best shot since Eve doesn't possess the original signal as the ground truth. The highlighted red dots in (a) and (b) show the trails where the performance of Eve's best attack was closest to Bob's reconstruction. The corresponding time-domain waveforms are plotted Fig. \ref{exp_attack} (c). 
Eve's reconstruction didn't reveal meaningful information about the neural signals. Fig. \ref{exp_attack} (d) shows the distribution of Bob's reconstructions and Eve's attacks. The results suggest that the CS neural recorder is safe from CoA attacks.

\begin{figure}[!ht]
  \centering
  \includegraphics[width=0.5\textwidth]{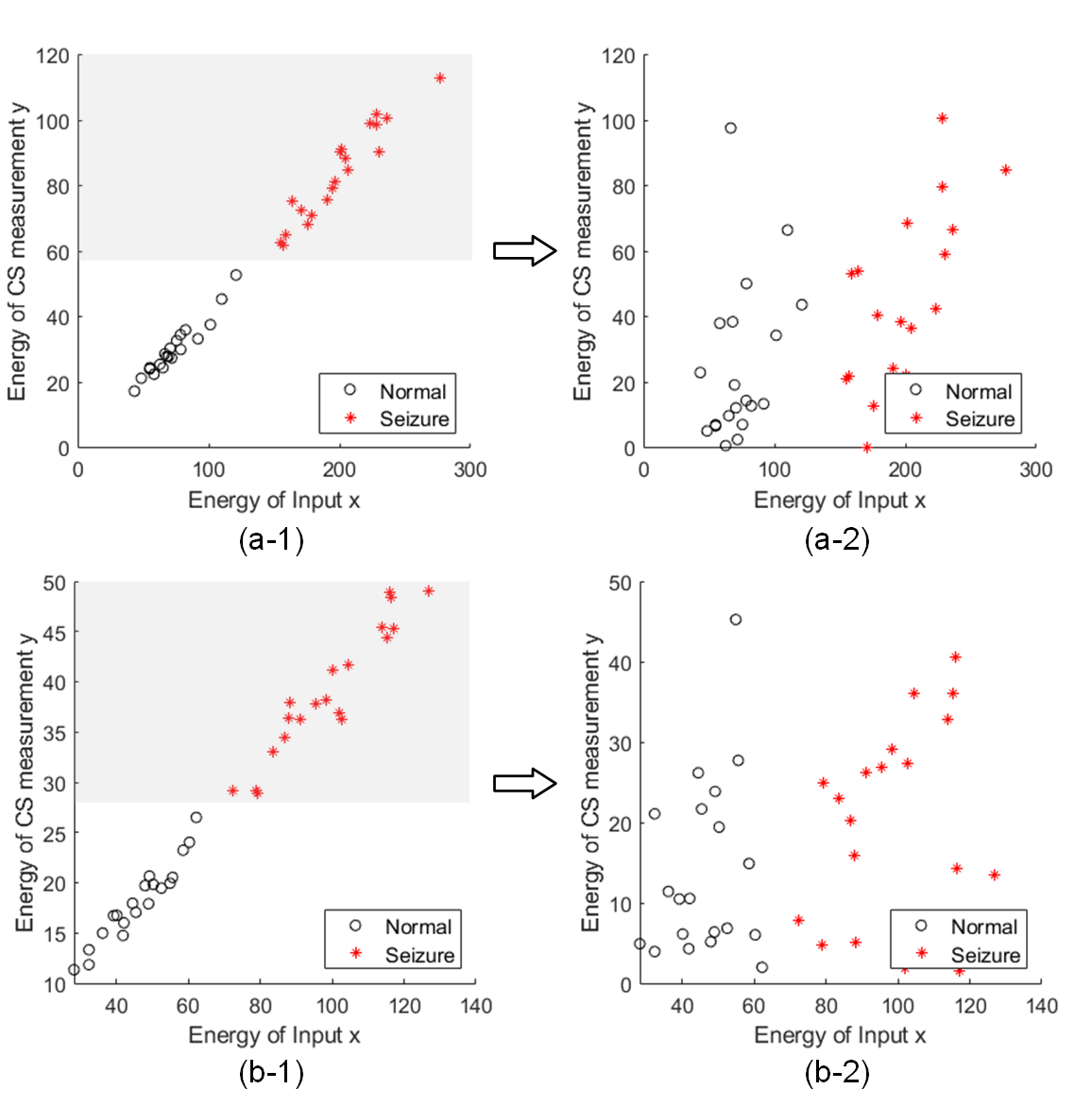}\\
  \caption{Experimental results of pseudo-random key shuffle. (a-1,2) and (b-1,2) show analysis of EEGs from two patients, with $x$-coordinate being the energy of the signal \textbf{x} and $y$-coordinate being the energy of the CS measurement \textbf{y}. Black circle markers show segments containing neural signals with normal activities, and red star markers show segments containing neural signals with seizure onset. The left column (a-1) and (b-1) use the same key for all CS measurements, and the right column (a-2) and (b-2) use the pseudo-randomly scrambled keys for the measurements of the same dataset.
  }\label{exp_key_scrambling}
\end{figure}

As discussed in Section \ref{sec2_cs}, the CS based encryption cannot prevent the energy of the signal \textbf{x} from being revealed, and the energy features of neural signals often contain valuable information. It should be noticed that the energy features of the neural signals are not the same as the energy of the wireless signals. To evaluate potential information leakage, we used the two patients' recordings with seizures. Long periods of interictal data segments were removed to speed up the experiment. Each marker in Fig. \ref{exp_key_scrambling} indicates a data segment with $x$-coordinate being the energy of the signal \textbf{x} and $y$-coordinate being the energy of the CS measurement \textbf{y} (without decipher). 
In addition, a circle marker (black) indicates the segment contains neural signals with normal activities, while a star marker (red) indicates the segment contains neural signals with seizure onset. The plots in the left column ((a-1) and (b-1)) use the same key for CS all measurements. The results show that seizure events can be classified using only the feature along the y-axis. In comparison, the plots in the right column ((a-2) and (b-2)) use the proposed key shuffle for the CS measurement of the same dataset. As expected, seizure classification is not possible using only the features in \textbf{y}. This experiment shows that the proposed scheme successfully places an additional layer of protection on the conventional CS based encryption.

As discussed in Section II.C, an adversary may non-invasively detect the power profile of the neural recording device and deduce timing characteristics of the encryption system from power analysis. Although we assumed that an adversary may acquire the timing information indirectly, we directly measured it during the experiments to evaluate the risk. Specifically, timing measurements were obtained using randomly generated key vectors. The measurement results confirmed that the execution of the ECC and ECDH protocols are time constant. No input-dependent branches were observed in 100,000 test runs. The results suggested that the scheme is safe against timing-based attacks.

\begin{figure}[!ht]
  \centering
  \includegraphics[width=0.48\textwidth]{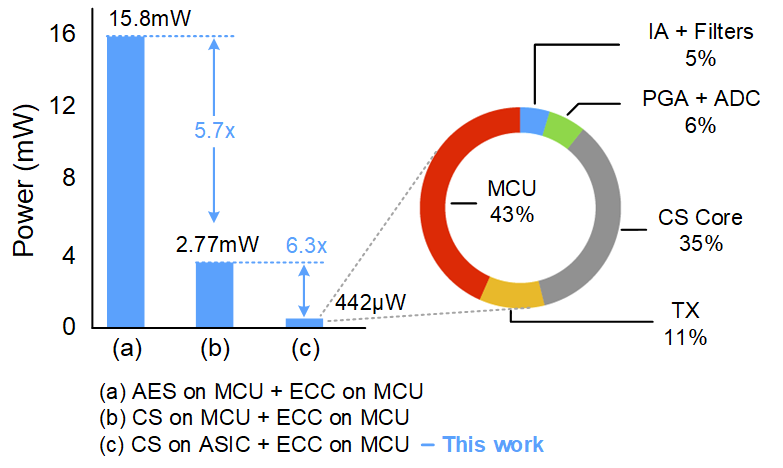}\\
  \caption{The measured power consumption of the developed wireless neural recording system in comparison with conventional implementations. The proposed encryption scheme achieved a 5.7x power saving for implementation all in MCUs, while the ASIC design further reduced the power by 6.3x. A 442$\mu$W was measured during a sampling rate of 500S/s and a CR of 8x. The detailed power breakdown is shown on the right.
  }\label{exp_power}
\end{figure}

The power consumption of the developed prototype was measured and compared with conventional implementations. Fig. \ref{exp_power} shows the measurement results with a detailed power breakdown.
To compare the result with conventional encryption schemes, we implemented a 256-bit AES on the MCU without data compression. The power consumption was measured to be 15.8mW including wireless transmission. Then we implemented the proposed ECC and CS hybrid scheme both on the MCU. The resulting power consumption was 2.77mW, corresponding to a 5.7x power saving. At last, the power consumption of the proposed device using the ASIC based CS and the MCU based key handling was 442$\mu$W. This was measured at a data rate of 500S/s and a CR of 8x. The results suggest that the developed prototype achieved an over 35x power saving compared with conventional encryption schemes.

The specifications and measured performance of the system are summarized in Table \ref{table_spec}. In addition, the energy efficiency for encryption is compared with prior low-power hardware encryption studies in Table \ref{table_power}. By taking advantage of the sparsity of neural signals, energy efficient ASIC design and MCU implementation, the proposed CS based encryption achieved a high energy efficiency.

\begin{table}[!ht]
\caption{Measured Specifications Summary}\label{table_spec}
\begin{tabular}{cc|cc}
\hline
\hline
Feature             & Value         & Feature     & Value      \\
\hline
ASIC   Technology   & 180nm         & CS Sampling    & 500S/s  \\
Supplies & 3.3V/1.8V & CS Ratio & 2x - 16x\\
IA gain             & 40 - 54dB     & ASIC clock   & up to 4MHz      \\
IA Noise            & 2.31$\mu$Vrms     & MCU clock  &32MHz    \\
IA Bandwidth        & 0.5 - 250Hz   & PSNR$^*$ & 32.75dB   \\
PGA Gain            & 4x - 7x       & $\rho^*$ & 0.973     \\
PGA Bandwidth       & 20kHz         & CS power    & 155$\mu$W  \\
ADC Rate & up to 40kSps  & Overall power       & 442$\mu$W      \\
ADC ENoB            & 9.3 bit       & Weight      & 4.7g      \\
\hline\hline
\end{tabular}
    \begin{threeparttable}
    \begin{tablenotes}
       \footnotesize
       \item $^*$  Reconstruction results from a CR of 8x.\\
     \end{tablenotes}
     \end{threeparttable}
\end{table}

\begin{table}[!ht]
\footnotesize
\caption{Comparison with Prior Low-Power Hardware Encryption Studies}
\label{table_power}
\begin{tabular}{cccccc}
\hline \hline
                 & \begin{tabular}[c]{@{}c@{}}2014 \\ \cite{sayilar2014cryptoraptor} \end{tabular}     & \begin{tabular}[c]{@{}c@{}}2014 \\ \cite{clercq2014ultra} \end{tabular}       & \begin{tabular}[c]{@{}c@{}}2018 \\ \cite{banerjee2018energy}\end{tabular}     & \begin{tabular}[c]{@{}c@{}}2018 \\ \cite{zhang2018recryptor}\end{tabular}    & \begin{tabular}[c]{@{}c@{}}This \\ work\end{tabular}                                                    \\ \hline
Hardware   & ASIC   & Cortex M0 & ASIC    & ASIC    & \begin{tabular}[c]{@{}c@{}}ASIC + \\ Cortex M0\end{tabular} \\ \hline
\begin{tabular}[c]{@{}c@{}}Data \\ compression \end{tabular} & No     & No        & No      & No      & 8x CS                                                        \\ \hline
\begin{tabular}[c]{@{}c@{}}Wireless \\ channel \end{tabular}         & No     & No        & No      & No      & Yes                                                         \\ \hline
\begin{tabular}[c]{@{}c@{}}Encryption \\ method \end{tabular}       & AES    & ECC       & SHA-2   & SHA-3   & CS + ECC                                                    \\ \hline
\begin{tabular}[c]{@{}c@{}}Energy \\ (norm.)\end{tabular}    & 124 nJ & 45.9 nJ   & 24.3 nJ & 48.7 nJ & 36.2 nJ$^*$                                                     \\ \hline \hline
\end{tabular}
    \begin{threeparttable}
    \begin{tablenotes}
       \footnotesize
       \item $^*$  IA, filters and wireless power is excluded for comparison.\\
     \end{tablenotes}
     \end{threeparttable}
\end{table}

\section{Discussion}
There are several limitations of this work that could be addressed with future research. First, authentication is important for exchanging public keys. Although the initial authentication of medical devices can often be conducted in a secure environment, such as during clinical visits, an integrated digital signature algorithm would improve the robustness and flexibility of the device. Established algorithms, such as the elliptic curve digital signature algorithm (ECDSA), can be implemented in the MCU \cite{Johnson2001}. Since the authentication process happens at a low frequency, it would not significantly impact the overall system power consumption.

Second, the dimension and resolution of the sampling matrices are important for achieving the optimal performance in terms of the reconstructed signal quality, maximum CR, security level, as well as hardware costs. The design trade-offs also include the targeted signal characteristics and the signal-to-noise ratio. It would be worth studying these trade-offs and implementing a configurable design in the future.

Third, differential power analysis (DPA) was not conducted in this work. DPA based attacks try to obtain the private keys by statistically analyzing the power consumption of the device \cite{Coron1999}. A thorough analysis and an IC level design that eliminates the risks from DPA would be an important step forward.

Finally, the MCU core can be integrated on-chip to further reduce the device form-factor and the power overhead (at an additional cost of silicon area). This is possible by integrating the Cortex-M0 core (freely available for research purpose \cite{arm2015}) or other open-source RISC-V processors. A wireless receiver (Rx) can be integrated on-chip for duplex wireless handshaking, so that the 2.4GHz transceiver can be removed from the system.

\section{Conclusion}

In this paper, we developed an energy-efficient wireless neural recording system with simultaneous data compression and encryption. The system integrated an ultra-low power CS ASIC and a general-purpose MCU. Novel techniques have been proposed to eliminate the risks from malicious attacks while maintaining an ultra-low power consumption. Experimental results showed that the developed system achieves a secure, reliable, energy-efficient neural recording over time.
Data encryption technology will be needed as therapies involving wireless neural interfaces become more prevalent in the treatment of neurological disorders \cite{naufel2020darpa}. Moreover, the scheme and circuit techniques introduced in this paper can be applied to a wide range of applications where high energy-efficiency and security are required.

\ifCLASSOPTIONcaptionsoff
  \newpage
\fi


\end{document}